\documentclass[aps,prl,twocolumn,showpacs,citeautoscript]{revtex4-1}
\usepackage{graphicx}
\usepackage{bm,amsmath,amssymb,mathrsfs}
\usepackage{hyperref}
\usepackage[usenames]{color}
\usepackage{dcolumn}
\usepackage{subfigure}
\usepackage{units}
\hypersetup{pdfborder=0 0 0,colorlinks=true,citecolor=blue,linkcolor=blue}
\input epsf

\begin{document}

\title{Band gaps in incommensurable graphene on hexagonal boron nitride}

\author{Menno Bokdam}
\author{Taher Amlaki}
\author{Geert Brocks}
\author{Paul J. Kelly}
\affiliation{Faculty of Science and Technology and MESA$^{+}$ Institute for
Nanotechnology, University of Twente, P.O. Box 217, 7500 AE Enschede, The
Netherlands}

\begin{abstract}
Devising ways of opening a band gap in graphene to make charge-carrier masses finite is essential for many applications. Recent experiments with graphene on hexagonal boron nitride ($h$-BN) offer tantalizing hints that the weak interaction with the substrate is sufficient to open a gap, in contradiction of earlier findings. Using many-body perturbation theory, we find that the small observed gap is what remains after a much larger underlying quasiparticle gap is suppressed by incommensurability. The sensitivity of this suppression to a small modulation of the distance separating graphene from the substrate suggests ways of exposing the larger underlying gap. 

%The effects of incommensurability and orientational misalignment on the electronic structure of graphene on a hexagonal boron nitride ($h$-BN) substrate are studied using tight-binding Hamiltonians parameterized to fit first-principles calculations.
%At the level of density functional theory (DFT), the band gap that is opened in commensurable graphene$|h$-BN by the sublattice-symmetry-breaking interaction with $h$-BN is quenched to $\leq 5$~meV by the 1.8\% lattice mismatch. 
%The band gap is greatly enhanced when the Coulomb interaction is taken into account in the GW approximation of many-body perturbation theory and a local height modulation is allowed.
%It is found to be $\sim 30$~meV when the lattices are aligned and to be quenched by a small misalignment, consistent with recent experimental results. 
\end{abstract}

\date{\today}

%\pacs{73.63.-b, 73.40.-c, 73.22.Pr, 73.20.-r, 73.20.At, 81.05.U-}

\maketitle

\paragraph{\color{red} Introduction.} 

The constant velocity of the charge carriers in graphene that results from the linear dispersion of the energy bands about the Dirac point gives rise to many of its intriguing properties \cite{Geim:natm07,Geim:sc09,Neto:rmp09} but also poses a serious limitation to its application in high-performance electronic devices \cite{Novoselov:nat12,Schwierz:pieee13}. For logic applications, transistors with on-off ratios of order $10^6$ are needed, requiring band gaps of $\sim 0.4$~eV \cite{Schwierz:pieee13}. Different approaches have been adopted to open a gap, the most promising of which is to use the interaction of graphene with a substrate to modify the linear dispersion of the bands. 

The current front runners to open a band gap using a sublattice 
symmetry-breaking interaction are SiC 
\cite{Mattausch:prl07,*Varchon:prl07,*Zhou:natm07} and hexagonal boron nitride 
($h$-BN) substrates  \cite{Giovannetti:prb07,Dean:natn10}. Because of its 
flatness, similarity to graphene, and the development of practical methods for 
preparing single layers of graphene on hexagonal boron nitride substrates 
\cite{Dean:natn10} there has been an explosion in the number of studies of this 
system. $h$-BN is a very suitable insulating substrate for making graphene-based 
devices \cite{Britnell:sci12} because it has dielectric characteristics similar 
to those of SiO$_2$, but contains fewer charged impurities and is atomically 
flat \cite{Lee:apl11,Britnell:nanol12}. These properties result in a higher 
charge carrier mobility for graphene on $h$-BN compared to graphene on SiO$_2$ 
\cite{Dean:natn10}, and in electron-hole puddles \cite{Martin:natp08} that are 
larger in size and less deep \cite{Xue:natm11,Decker:nanol11}. 

In this paper we will show that the recently observed gap of order 
30~meV \cite{Hunt:sc13} results
from a large many-body enhanced quasiparticle gap being canceled by the 
incommensurability of graphene and $h$-BN that is only partially restored by a 
lateral variation of the height of graphene above the $h$-BN substrate. The 
large size of the underlying band gap and the mechanism of its cancellation 
suggest ways of recovering the large bare bandgap.  

Graphene on top of $h$-BN experiences a perturbing potential comprising two 
components. First, the 1.8\% lattice mismatch between the two honeycomb 
lattices and orientational misalignment give rise to a slowly varying component 
that has been observed as moir\'e patterns in scanning tunneling microscopy 
images \cite{Xue:natm11,Decker:nanol11}. 
Because of the chiral nature of the states close to the Dirac point \cite{Ando:jpsj98a,*Ando:jpsj98b}, this component does not open a gap \cite{Park:natp08,Barbier:prb08,Park:prl08,Burset:prb11}; the superlattice Dirac points predicted by these effective Hamiltonian theoretical studies have recently been observed in scanning tunneling spectroscopy experiments \cite{Yankowitz:natp12}.

Second, the heteropolar $h$-BN substrate also gives rise to a sublattice 
symmetry-breaking potential that opens a gap at the Dirac point when graphene 
and $h$-BN are commensurable \cite{Giovannetti:prb07}.
The initial failure \cite{Dean:natn10,Xue:natm11,Decker:nanol11} to observe the band gap of order 50 meV predicted by these first principles calculations was attributed to the lattice mismatch \cite{Xue:natm11,Sachs:prb11}. On the basis of binding energies calculated from first principles, it was argued that the energy gained by graphene bonding commensurably to $h$-BN was insufficient to offset the energy cost of achieving this by stretching graphene (or compressing $h$-BN). Tight-binding (TB) analyses led Xue \cite{Xue:natm11} and Sachs \cite{Sachs:prb11} to argue that the symmetry-breaking interaction between graphene and $h$-BN would average out in the incommensurable case to a much smaller (but not vanishing \cite{Kindermann:prb12}) band gap.

Recent temperature-dependent transport studies indicating the occurrence of a 
metal-to-insulator transition at the charge neutrality point at low 
temperatures, suggest that the situation may be more complex 
\cite{Ponomarenko:natp11,Amet:prl13,Hunt:sc13}.  The low temperature insulating 
state has been interpreted to be induced by disorder (Anderson transition) 
\cite{Ponomarenko:natp11} or to result from substrate-induced valley symmetry 
breaking \cite{Amet:prl13,Hunt:sc13}. It has also been suggested that even small 
symmetry-breaking-induced band gaps may be greatly enhanced by many-body 
interactions \cite{Song:arxiv13}.

Because the gap suppression by lattice mismatch depends on details of the 
approximations made in deriving effective Hamiltonians for graphene 
\cite{Kindermann:prb12}, we use first-principles calculations to derive explicit 
$\pi$-orbital nonorthogonal TB Hamiltonians for graphene on $h$-BN that do not 
appeal to perturbation theory to fold the graphene-substrate interaction into an 
effective Hamiltonian. The band gap induced by a commensurable $h$-BN substrate 
survives the small lattice mismatch between graphene and $h$-BN but is indeed 
greatly reduced; it does not survive misaligning the two lattices.  
%Using the LDA eigenvalues, the maximum size of the gap is 5~meV. 
Taking into account a local variation of the graphene-$h$-BN separation and 
uqasiparticle energies calculated within the \textit{GW} approximation of 
many body perturbation theory leads to a band gap of 32~meV.

\paragraph{\color{red} Computational details.}

We use density functional theory (DFT) at the level of the local density 
approximation (LDA) \cite{Perdew:prb81} within the framework of the plane-wave 
projector augmented wave (PAW) method \cite{Blochl:prb94b}, as implemented in 
VASP \cite{Kresse:prb93,Kresse:prb96,Kresse:prb99}, to determine the 
electronic structure of a graphene sheet on top of a $h$-BN substrate. 
%as a function of the height $z$ and lateral displacement $x,y$ of the two parallel lattices. 
A plane wave basis with a cutoff energy of 400 eV is used in combination with a 
$36 \times 36$ $k$-point grid (in a $1\times 1$ unit cell).
Many-body effects are studied within the GW approximation \cite{Hedin:pr65} 
starting with LDA Kohn-Sham (KS) orbitals \cite{Hybertsen:prb86} calculated for 
a cell containing two $h$-BN layers and a graphene layer. We use the $GW$ 
implementation in VASP \cite{Shishkin:prb06}, with 12 occupied and 52 empty 
bands and 50 points on the frequency grid. Interactions between periodic images 
in the $z$ direction lead to a dependence of the $GW$ band gap on the cell size 
that we remove by linearly extrapolating the calculated gaps as a function of 
the inverse cell size to infinite separation \cite{Berseneva:prb13}.

\paragraph{\color{red} Density functional calculations.}

\begin{figure}
\includegraphics[scale=0.61]{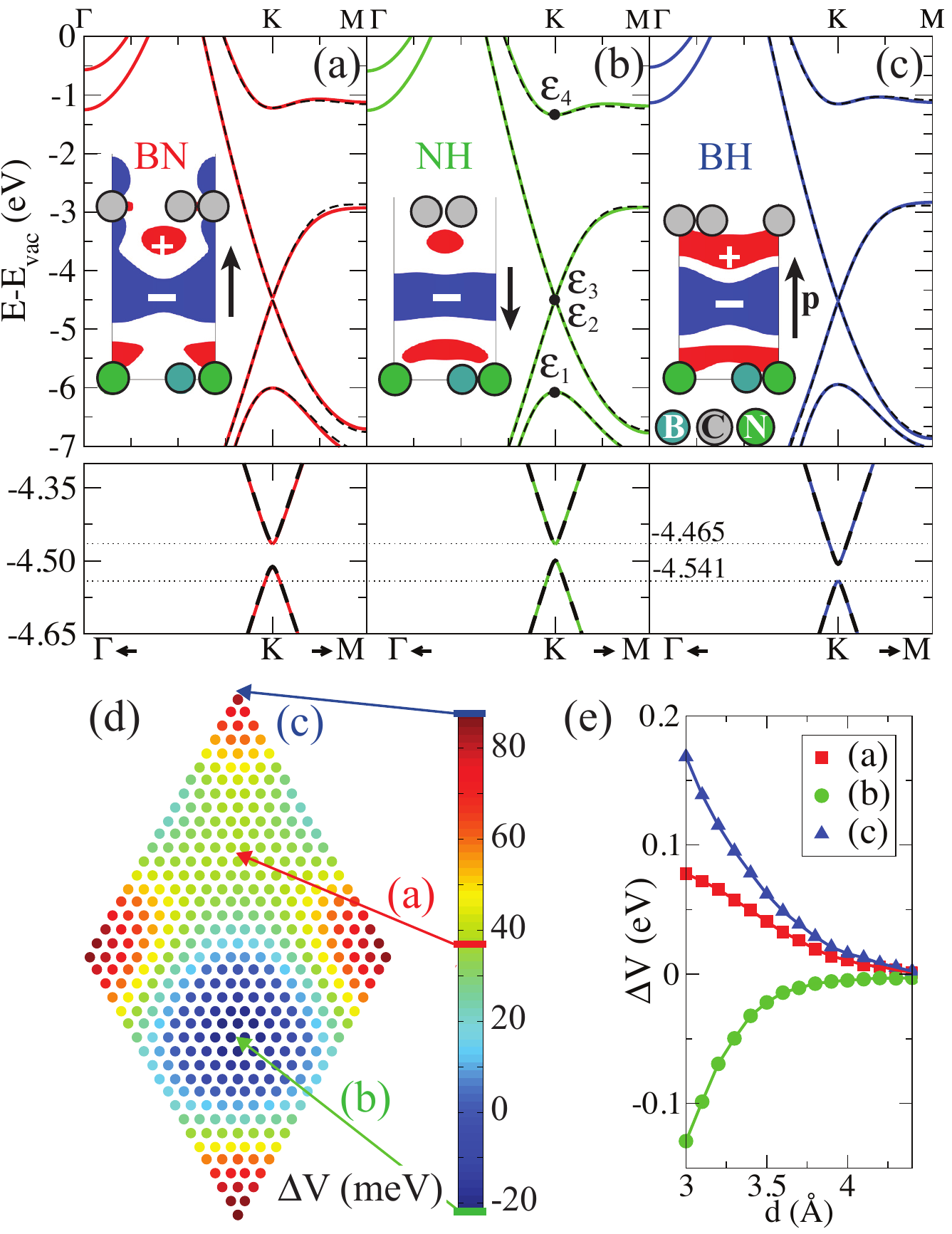}
\caption{(Color) LDA band structures of graphene on $h$-BN for the (a), (b), and 
(c) commensurable structures, with respect to a common vacuum level. Solid and 
dashed lines give the DFT results and the TB fits, respectively. The region 
around the Dirac point is enlarged in the panels below. The formation of an 
interface dipole is illustrated by the charge displacement $-e\Delta n$ in the 
$yz$ plane containing B, C and N atoms. Blue and red indicate regions of 
negative and positive charge density, respectively, giving a dipole moment 
\textbf{p}. (d) The interface potential step $\Delta V$ as a function of the 
position $(x,y)$ of  graphene with respect to $h$-BN, at the RPA equilibrium 
separation $d_\mathrm{eq}(x,y)$ [see Fig.~\ref{figB}a]. (e)  Dependence of 
$\Delta V$ on $d$ for configurations (a)-(c).}
\label{figA}
\end{figure}

We start by analyzing the electronic structures of rotationally aligned, 
commensurable graphene$|h$-BN  at the DFT and $GW$ levels, before using the 
results to construct a TB Hamiltonian for rotated, incommensurable structures. 
We focus for convenience on the three high symmetry configurations: (a) with one 
carbon over B, the other over N; (b) with one carbon over N, the other centered 
above a $h$-BN hexagon; and (c) with one carbon over B, the other centered 
above a $h$-BN hexagon \cite{Giovannetti:prb07}. Panels (a)-(c) of 
Fig.~\ref{figA} show the LDA energy bands for graphene on $h$-BN for the (a), 
(b) and (c) configurations at their RPA equilibrium separations of $d_{\rm eq} 
=$ 3.55, 3.50 and 3.35\,\AA{} \cite{Sachs:prb11}, respectively, with LDA gaps 
$\Delta \varepsilon \equiv  \varepsilon_{3}(K)-\varepsilon_{2}(K)$ of 
30-40 meV opened at the $K$ point by the symmetry-breaking interaction with the 
substrate \cite{Giovannetti:prb07}. The Dirac point is situated asymmetrically 
in the $h$-BN band gap about a third of the way from the top of the valence 
band. Apart from the formation of the small gap,
%$\Delta \varepsilon$, 
the dispersion of the $\pi$ bands is largely unchanged by the interaction with the $h$-BN substrate within about 1~eV of the Dirac point. 

If we expand the energy scale about the Dirac point, we see that the centers of the gaps, $[\varepsilon_2({\rm K}) + \varepsilon_3({\rm K})]/2$, do not coincide for the different configurations (lower panels); the interaction between graphene and $h$-BN gives rise to an interface potential step $\Delta V$. 
The source of this potential step is a configuration-dependent interface dipole layer that can be visualized in terms of the electronic displacement $\Delta n({\bf r})$ obtained by subtracting the electron densities of the isolated constituent materials, $n_{\rm Gr}$ and $n_{\rm BN}$, from that of the entire system $n_{\rm Gr|BN}$ \cite{Bokdam:nanol11,*Bokdam:prb13}. The potential step is related to the dipole layer, illustrated in Fig.~\ref{figA} for the (a), (b) and (c) configurations, by $\Delta V= -e^2/ (A\epsilon_0) \int z \Delta n({\bf r}) \,dx\, dy\, dz $ where $A$ is the area of the surface unit cell and the integration is over all space.

The potential step depends sensitively on how the graphene and $h$-BN lattices are positioned. Starting from the (c) configuration, and displacing graphene laterally by $(x,y)$ yields the potential landscape $\Delta V(x,y)$ shown in Fig.~\ref{figA}(d). 
For each value of $(x,y)$, the graphene sheet is a distance 
$d_\mathrm{eq}(x,y)$, the RPA equilibrium separation, from the $h$-BN 
substrate.  $\Delta V$ reaches appreciable values ranging from $-20$ to $90$ 
meV, where the minimum and maximum values are found for the high symmetry 
configurations (b) and (c), respectively. 
The full $d$-dependence of $\Delta V$ is shown for the three configurations in Fig.~\ref{figA}(e).

\paragraph{\color{red} GW correction}

\begin{figure}[t]
\includegraphics[width=8.5cm]{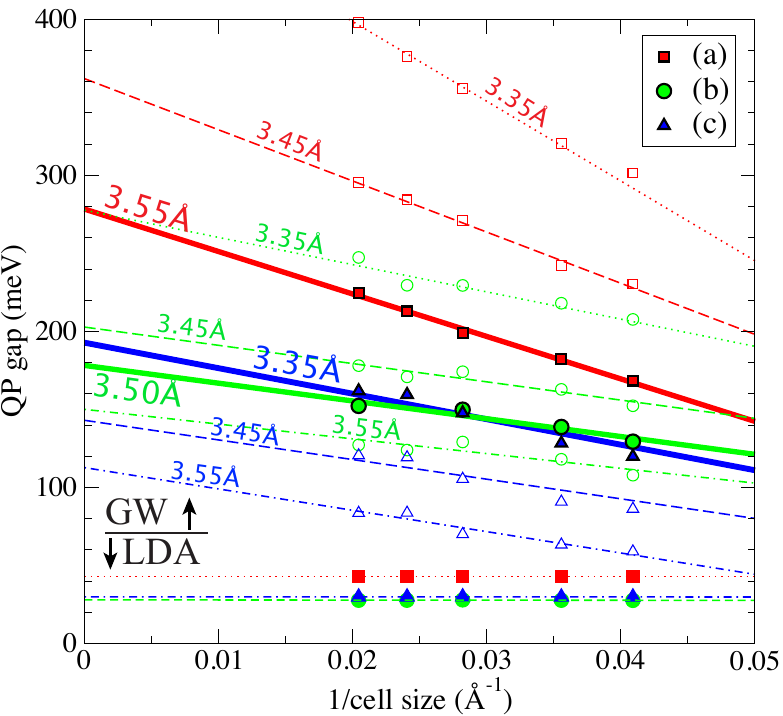}
\caption{(Color) Quasiparticle (QP) and LDA gaps opened at the Dirac point as a 
function of the inverse cell size. The thick solid lines correspond to graphene 
at the RPA equilibrium distances of configurations (a), (b) and (c) 
\cite{Sachs:prb11}. The dotted, dashed and dashed/dotted lines correspond to 
graphene$-h$-BN separations of 3.35, 3.45, and 3.55 \AA{}, respectively.} 
\label{figC}
\end{figure}

We perform $G_0W_0$ calculations for the three symmetric configurations in 
order to find the many body corrections to the LDA band gaps. Figure~\ref{figC} 
shows the quasiparticle gaps as a function of the inverse cell size $1/a$ in the 
direction perpendicular to the graphene sheet \cite{fn:CT} for the three 
symmetric configurations and for values of the height of graphene above the 
$h$-BN substrate that span the full range of RPA equilibrium separations. We are 
interested in the band gap for an isolated layer of graphene on $h$-BN that we 
determine by extrapolation to $a \rightarrow \infty$. The gap is seen to 
increase dramatically, from 43, 28, and 30 meV (horizontal lines) in the LDA to 
278, 178 and 193 meV, for the (a), (b) and (c) configurations at the RPA 
equilibrium separations (thick lines), respectively.

\paragraph{\color{red} Band gap in graphene on $h$-BN}

Before we use the results of the full $ab initio$ calculations for the 
commensurable systems to construct a TB Hamiltonian for the incommensurable 
case, it is useful to estimate the size of the band gaps we expect to find. We 
can make use of the weakness of the interaction between the graphene and $h$-BN 
and the large separation in energy between the Dirac point eigenvalues and the 
$h$-BN valence and conduction band edges to construct an effective Hamiltonian 
for the graphene layer by L\"{o}wdin downfolding, otherwise know as the Schur 
complement. The substrate can then be replaced by three effective potentials 
\cite{Kindermann:prb12}: one that locally shifts the Dirac point, another that 
opens a local gap, and a third that has the form of a pseudo-magnetic field. By 
applying first order degenerate perturbation theory to the downfolded 
Hamiltonian, we estimate the gap of the incommensurable system to be 
\begin{equation}
\label{eq:avgap}
\overline{\Delta \varepsilon}= \left|\Delta \varepsilon_{\rm (b)}+\Delta 
\varepsilon_{\rm (c)}-\Delta \varepsilon_{\rm (a)} \right| / 3,
\end{equation}
in terms of the gaps of the commensurable systems. 

This estimate should be valid in the limit that the mismatch is very small and all possible configurations are sampled equally. 
Using Eq.~\eqref{eq:avgap} and assuming that the structure is locally at its RPA 
equilibrium separation, we expect the band gap of aligned, incommensurable 
graphene on $h$-BN to be $31$ meV at the $G_0W_0$ level, as compared to 5 meV 
at the LDA level. Assuming the graphene sheet is flat results in much smaller 
$G_0W_0$ gaps of 10, 5 and 4 meV for separations of 3.35, 3.45, and 3.55 \AA{}, 
respectively. This indicates that it is important to take the modulation of the 
equilibrium separation into account. We will find that the gaps estimated using 
Eq.~\eqref{eq:avgap} are very close to the results obtained by full numerical 
diagonalization of the TB Hamiltonian. 

\paragraph{\color{red} Tight-binding Hamiltonian.}

The diagonal blocks of the TB Hamiltonian and overlap matrices
\begin{equation}
H = \left(\begin{array}{cc}  H_{\rm Gr}               & H_{\rm int}   \\ 
                             H_{\rm int}^{\dagger}    & H_{\rm BN}    \\
            \end{array} \right) \; ; \;
O = \left(\begin{array}{cc}  O_{\rm Gr}               & O_{\rm int}   \\ 
                             O_{\rm int}^{\dagger}    & O_{\rm BN}    \\
           \end{array} \right)      
\label{eq:TBH}
\end{equation}
are first determined separately for monolayers of graphene and $h$-BN. Because 
isolated monolayers have reflection symmetry, the $p_z$ blocks of the 
corresponding matrices are decoupled from the $\{s, p_x,p_y \}$ blocks. By 
introducing an overlap matrix, $H$ and $O$ can be chosen to have short range and 
the $ab initio$ (LDA and $GW$) bands can be fit essentially exactly. When we 
consider the interaction between graphene and $h$-BN, we include the interface 
potential step $\Delta V$ as an additional diagonal term. 

\begin{figure}
\includegraphics[scale=0.4]{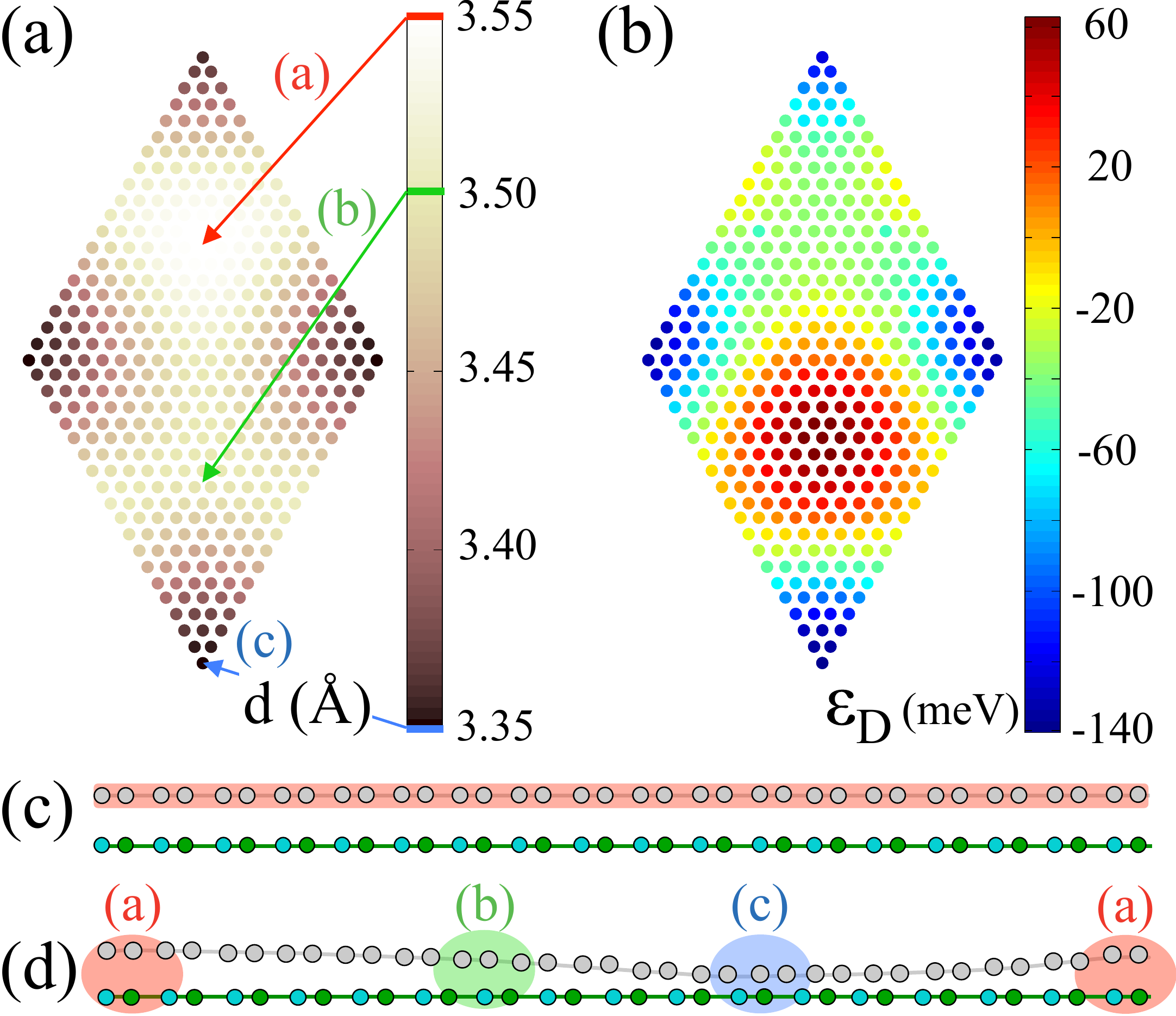}
\caption{(Color) (a) The RPA equilibrium separation as a function of the 
lateral displacement of the two lattices for commensurable graphene on $h$-BN. 
(b) The position of the Dirac point  with respect to the center of the $h$-BN 
band gap, $\varepsilon_{\rm D}$. The plotted values are relative to the 
constant $\varepsilon_{\rm D}$ obtained for graphene and $h$-BN at an infinite 
interlayer distance. Schematic representations of (c) commensurable and (d) 
incommensurable configurations of graphene on $h$-BN where the bonding locally 
resembles commensurable configurations.}
\label{figB}
\end{figure}

We determine $\Delta V$ for incommensurable structures by interpolation and approximate incommensurable lattices by periodic superstructures. For example, the factor $1.018$ between the $h$-BN and the graphene lattice parameters can be represented by the rational approximant $56/55$, which leads to a supercell of $56 \times 56\times 2$ carbon atoms and $55 \times 55$ each of boron and nitrogen atoms. As the lattice mismatch is small, we assume that the graphene sheet locally follows the RPA equilibrium separation $d_\mathrm{eq}(x,y)$ as a function of position $(x,y)$ within this supercell, as shown in Fig.~\ref{figB}(a). The difference between the commensurable and incommensurable structure is schematically illustrated in Figs.~\ref{figB}(c) and \ref{figB}(d). The potential step $\Delta V(x,y)$ can then be obtained from Fig.~\ref{figA}(d). 

To define $\Delta V$ when the graphene lattice is rotated through an angle $\phi$ with respect to the $h$-BN substrate, we make use of the fact that $\Delta V(x,y) $ has threefold rotation symmetry and interpolate for intermediate angles. This approximation was checked for commensurable lattices where explicit DFT calculations can be performed for specific rotation angles for which the unit cell sizes are manageable.

\paragraph{\color{red} TB: Commensurable lattices.}

The remaining parameters of the TB Hamiltonian are obtained from fits to either 
the LDA or the GW band structures of commensurable graphene$|h$-BN structures. 
The band structures calculated with the TB Hamiltonian are compared to the LDA 
results in Figs.~\ref{figA}(a)-(c) (dashed lines). The TB Hamiltonian clearly 
yields a satisfactory fit for the three configurations shown. It also accurately 
captures the shift of the Dirac cone as a function of an $xy$ translation of the 
graphene sheet over the $h$-BN substrate. Figure \ref{figB}(b) shows the 
position of the Dirac point (the center of the gap) with respect to the center 
of the $h$-BN band gap,
\begin{equation}
\varepsilon_{\rm D}(x,y)=
%\nicefrac{1}{2}
\left\lbrace [\varepsilon_3({\rm K})+\varepsilon_2({\rm K}) ] -
             [\varepsilon_4({\rm K})+\varepsilon_1({\rm K}) ] \right \rbrace /2,
\end{equation}
for the height profile shown in Fig.~\ref{figB}(a); it fits the LDA results 
exactly. Comparison of the LDA energy bands for a structure with graphene 
rotated through $21.78^\circ$ with respect to $h$-BN (resulting in a $\sqrt{7} 
\times \sqrt{7}$ unit cell with 14 carbon, 7 boron and 7 nitrogen atoms) again 
shows very good agreement. Because the rotated system was not used in fitting 
the TB Hamiltonian, the good description of the bands close to the Dirac points 
is an important measure of its predictive capability. 

\paragraph{\color{red}  TB: Incommensurable lattices.}

It has been argued that the bonding between graphene and a $h$-BN substrate is 
so weak and those honeycomb structures so stiff that the small 1.8\% lattice 
mismatch will persist in graphene$|h$-BN structures \cite{Sachs:prb11}. For 
perfect alignment, i.e., $\phi=0$, this implies that even if we start at a 
position where the local bonding corresponds to the  lowest energy [the (c) 
bonding configuration], the lattice mismatch will result in dephasing of the 
two lattices. However, the mismatch is sufficiently small that locally the 
hopping is indistinguishable from a commensurable system displaced by some 
amount $(x,y)$, with an equilibrium distance $d_\mathrm{eq}(x,y)$, so the 
Hamiltonian for the incommensurable system can be assembled with the 
parametrization described above. Choosing the vacuum potential as the common 
potential zero leads to diagonal elements of the TB Hamiltonian that depend on 
the local interface potential $\Delta V(x,y)$. The $h$-BN conduction and valence 
band edges then undulate in real space, a prediction that could be confirmed by 
experiment.

When $\phi=0$ and assuming that the graphene sheet follows the height profile of 
Fig.~\ref{figB}(a), we calculate a ($GW$) band gap of 32~meV. This is 
consistent with reports of the charge neutrality point resistance increasing 
with decreasing temperature \cite{Ponomarenko:natp11,Amet:prl13} and of gaps 
between 16 and 28~meV that were extracted from temperature dependent 
resistivity measurements \cite{Hunt:sc13}. The height variation in the structure 
turns out to be important; using a flat graphene sheet at separations of 3.35, 
3.45, and 3.55 \AA{}, results in gaps of 10, 6, and 5 meV, respectively. This 
suggests that it may be possible to realize larger gaps in graphene by suitably 
modulating the graphene-$h$-BN spacing.
For rotation angles $\phi>0^\circ$, the gap rapidly becomes smaller. Calculations for a $\phi=6^\circ$, $\delta=2.0$\%{} incommensurable structure result in a vanishing gap. Further studies on even larger supercells corresponding to smaller angles have to be done to determine the critical angle at which a band gap is opened.

\paragraph{\color{red} Conclusions.}

Many-body calculations for commensurable graphene on $h$-BN show that 
quasiparticle band gaps are much larger than previously thought. 
Incommensurability results in a near complete cancellation of these gaps, but a 
detailed analysis suggests that it may be possible to realize gaps much larger 
than the recently observed 30~meV by suitably modulating the separation of 
graphene from the substrate. This might be achieved by structuring the 
substrate, by exciting phonons with wavelengths that frustrate the cancellation 
of the gap-opening interactions, or by applying modest pressure. The Hamiltonian we have derived without making recourse to 
perturbation theory \cite{[A procedure similar to that presented here has been developed by ][ 
; albeit at the LDA level i.e. without considering many-body effects 
]Jung:arXiv13,*[and used to explain the bandgap in terms of bond distortions 
alone ]Jung:arXiv14} 
can be used to reexamine the validity of phenomenological studies of how a h-BN 
substrate affects the electronic structure of graphene 
\cite{Ortix:prb12,Wallbank:prb13a,*Wallbank:prb13b} where the interaction was 
considered perturbatively.  

%The substrate interaction for rotationally aligned graphene on $h$-BN opens up a gap at the Dirac point of graphene that survives the 1.8\%{} lattice mismatch. We calculate a gap of 5 meV at the LDA level, which is enhanced by many body interactions to 32 meV at the GW level, in agreement with recent experiments \cite{Hunt:sc13}. The variation in the graphene$|h$-BN distance ($\pm$0.1 \AA{}) along the interface is important; at a fixed distance the gap would be $\lesssim 10$ meV. 

\acknowledgements{This work is part of the research program of the Foundation 
for Fundamental Research on Matter (FOM), which is part of the Netherlands 
Organisation for Scientific Research (NWO).  The use of supercomputer facilities 
was sponsored by the Physical Sciences division of NWO (NWO-EW). M.B. 
acknowledges support from the European project MINOTOR, Grant No. 
FP7-NMP-228424.}


\begin{thebibliography}{47}
\bibitem [{fn:()}]{fn:CT}%
  \BibitemOpen
  \href@noop {} {}\bibinfo {note} {For periodically repeated slabs, the images
  screen the excitations. For sufficiently large separations, the image
  potential effect decays as $1/a$ and we determine the gap for an isolated
  graphene sheet on $h$-BN by extrapolating to $a = \infty, 1/a = 0$
  \cite{Berseneva:prb13,Huser:prb13}. Data points corresponding to the largest
  cell size were given the greatest weight in the weighted fitting and the
  average gap, Eq.~\eqref{eq:avgap}, was required to decrease monotonically as
  a function of the separation.}\BibitemShut {Stop}%
\end{thebibliography}
\end{document}